\documentclass[reprint,amsmath,amssymb,aps,prc,superscriptaddress]{revtex4}
\usepackage{lineno}
\usepackage{graphicx}
\usepackage{dcolumn}
\usepackage{bm}
\usepackage[utf8]{inputenc}
\usepackage{comment}
\usepackage{subcaption}
\usepackage{hyperref}
\usepackage{tikz}
\usepackage{array}
\usepackage{tabularx}
\usepackage{multirow}
\usepackage{cleveref}
\usepackage{booktabs}
\newcolumntype{C}{>{$}c<{$}}

\AtBeginDocument{%
  \heavyrulewidth=.08em
  \lightrulewidth=.05em
  \cmidrulewidth=.03em
  \belowrulesep=.65ex
  \belowbottomsep=0pt
  \aboverulesep=.4ex
  \abovetopsep=0pt
  \cmidrulesep=\doublerulesep
  \cmidrulekern=.5em
  \defaultaddspace=.5em
}

\usepackage{xcolor}
\usepackage{ulem}
\setcitestyle{maxcitenames=1}
\newcolumntype{P}[1]{>{\centering\arraybackslash}p{#1}}
\newcolumntype{M}[1]{>{\centering\arraybackslash}m{#1}}
\usepackage{pdfpages}

\hypersetup{%
    colorlinks=true
    ,urlcolor=blue
    ,citecolor=blue
    ,linkcolor=blue
    }

\usepackage{caption}
\begin{document}

\title{Proton Number Fluctuations in Au+Au and Ag+Ag Collisions at \(\sqrt{s_{\mathrm{NN}}}=2.4\ \mathrm{GeV}\): A JAM Model Study}
\author{Yongcong Xu}
\author{Zhaohui Wang}
\affiliation{Key Laboratory of Quark \& Lepton Physics (MOE) and Institute of Particle Physics, \\ Central China Normal University}
\author{Yu Zhang}
\affiliation{School of Physical Science and Technology \& Guangxi Key Laboratory of Nuclear Physics and Technology, Guangxi Normal University}
\author{Xiaofeng Luo } \thanks{xfluo@ccnu.edu.cn} 
\affiliation{Key Laboratory of Quark \& Lepton Physics (MOE) and Institute of Particle Physics, \\ Central China Normal University}

\date{\today}
\begin{abstract}

Proton number fluctuations in low-energy heavy-ion collisions are sensitive to 
nuclear interactions, baryon-number conservation, and fluctuations of the collision volume, 
all of which can complicate searches for critical phenomena. We investigate proton-number 
cumulants and factorial cumulants up to fourth order in Au+Au and Ag+Ag collisions at 
\(\sqrt{s_{\mathrm{NN}}}=2.4\ \mathrm{GeV}\) using the JAM hadronic transport model in cascade 
and mean-field modes. We employ the Centrality-Independent Genuine Cumulant Analysis fRamework 
(CIGAR) to suppress the initial-volume fluctuations and compare its results with 
those obtained using conventional centrality-bin-width corrections. CIGAR closely reproduces 
the participant-number-based reference, whereas the multiplicity-based analysis exhibits 
substantial deviations. In both collision systems, the ratios $C_{2}/C_{1}$, $C_{3}/C_{1}$, 
and $C_{4}/C_{2}$ generally decrease toward more central collisions and with increasing 
rapidity acceptance, while the mean field modifies their magnitudes. A simplified treatment 
of light-nucleus formation through probabilistic proton removal shifts $C_{4}/C_{2}$ toward 
less negative values. These results establish a hadronic-transport reference without an 
explicit QCD critical contribution and highlight the importance of controlling centrality 
fluctuations, acceptance, and light-nucleus formation when interpreting experimental measurements 
at high baryon density.
\end{abstract}
\maketitle

\section{Introduction}\label{sec:introduction}

Mapping the phase structure of strongly interacting matter is a central objective of 
relativistic heavy-ion physics. The QCD phase diagram is commonly expressed in terms of 
temperature and baryon chemical potential, and theoretical studies suggest the possible 
existence of a critical point at the endpoint of a first-order phase-transition line~\cite{Bzdak:2019pkr,
Braun-Munzinger:2026krf,Luo:2022mtp,Koch2026,Clarke2024,PhysRevD.110.094006,
PhysRevD.101.054032,PhysRevD.104.054022,GAO2021136584,Sorensen2024,Cai:2024eqa,Zhu:2025gxo, lu2026extractingfreezeoutconditionsbeam}. 
Higher-order cumulants of conserved quantities, including net-baryon number, 
net-electric charge, and net-strangeness, are especially promising observables because of 
their sensitivity to multi-particle correlations and, near a critical point, 
to the correlation length~\cite{Stephanov:2008qz,ASAKAWA2016299,PhysRevC.96.024910,Luo2017,Li:2018ygx}. 
Experimentally, higher-order cumulants of net-proton, net-charge, and net-kaon multiplicity distributions have been measured by the STAR Collaboration across the RHIC beam-energy scan (BES), including precision net-proton fluctuation results, providing the primary experimental motivation for systematic fluctuation studies at finite baryon chemical potential~\cite{PhysRevLett.105.022302, PhysRevLett.112.032302, PhysRevLett.113.092301, PhysRevLett.126.092301, PhysRevC.104.024902, STAR:2025zdq}.
Nonmonotonic collision-energy dependence has therefore 
been proposed as a possible signature of critical behavior~\cite{Li:2017ple,Fu:2023lcm}. Establishing reliable 
noncritical reference calculations is essential, particularly in the high-baryon-density 
region reached at low collision energies.

The HADES Collaboration has measured proton-number fluctuations in Au+Au collisions at 
\(\sqrt{s_{\mathrm{NN}}}=2.4\ \mathrm{GeV}\), providing a valuable experimental constraint 
at very high baryon chemical potential~\cite{hades_2p4}. At this energy, nuclear interactions and 
potentially the nuclear liquid-gas transition may influence fluctuation observables 
in addition to any effects associated with the QCD phase structure~\cite{recchi2026imprintsnuclearliquidgasphase}. Throughout this paper, 
\(C_n\) denotes the \(n\)th-order ordinary proton-number cumulant and \(\kappa_n\) the 
corresponding factorial cumulant. The HADES notation differs: the quantities denoted \(K_n\) 
and \(C_n\) in Ref.~\cite{hades_2p4} correspond to \(C_n\) and \(\kappa_n\), respectively, in the present work. 
HADES reported proton cumulants up to fourth order and observed an approximately power-law 
dependence of factorial cumulants on the mean proton number, 
\(\kappa_n\propto \langle N_p\rangle^n\), suggestive of substantial multi-proton correlations 
in momentum space~\cite{hades_2p4}.

At such low energies, however, measured proton-number fluctuations also reflect baryon 
stopping, resonance production and decay, hadronic rescattering, nuclear cluster formation, 
the conservation of baryon number, fluctuations in the collision volume, and experimental 
acceptance and efficiency~\cite{fp85-w4l2}. Separating these contributions from possible critical 
correlations requires a microscopic description of the underlying hadronic dynamics. The JAM 
transport model provides such a description and has been used to investigate the influence of 
rescattering, decays, centrality fluctuations, and deuteron formation on proton cumulants at 
low collision energies~\cite{2019JAM,PhysRevC.61.024901,
PhysRevC.105.014911,PhysRevC.101.034909,Chatterjee_2021}. 
These JAM-based studies are directly relevant to the cumulant and correlation-function measurements reported by STAR for net-proton, proton, and antiproton multiplicity distributions across BES energies, as well as to the cumulants and factorial cumulants measured at \(\sqrt{s_{\mathrm{NN}}}=3\ \mathrm{GeV}\)~\cite{PhysRevC.104.024902, 2018551, PhysRevC.107.024908}.
Extending these calculations to \(\sqrt{s_{\mathrm{NN}}}=2.4\ 
\mathrm{GeV}\) is particularly timely in view of the HADES results. This is further reinforced by STAR measurements of proton and net-proton cumulants down to \(\sqrt{s_{\mathrm{NN}}}=3\ \mathrm{GeV}\) and by broader experimental reviews of fluctuation results from the RHIC BES program, which together define the experimental landscape that transport-model baselines need to address~\cite{PhysRevC.107.024908, Chen2024}.

A further challenge is that conventional centrality definitions become less precise as the 
collision energy decreases. Because charged-particle multiplicities are relatively small, 
residual collision-volume fluctuations can significantly bias higher-order cumulants even 
after applying standard centrality-bin-width corrections. We therefore analyze the JAM events 
using the Centrality-Independent Genuine Cumulant Analysis fRamework (CIGAR)~\cite{WANG2025139984}. 
This approach reconstructs centrality-resolved proton-number distributions from an overall 
multiplicity distribution using an Edgeworth-based parameterization and an optimization 
procedure combining differential evolution with Bayesian inference. In the present study, 
we examine how well this method suppresses initial-volume-fluctuation biases at 
\(\sqrt{s_{\mathrm{NN}}}=2.4\ \mathrm{GeV}\).

We systematically compare Au+Au and Ag+Ag collisions in cascade and mean-field modes, 
focusing on the centrality and rapidity-acceptance dependence of ordinary and factorial 
cumulants. We also estimate, using a simplified proton-removal prescription, how the depletion 
of free protons by light-nucleus formation may modify these observables. Section~\ref{sec:jam_2p4} describes 
the JAM simulation and kinematic acceptance; Sec.~\ref{sec:moments} defines the fluctuation observables; 
Sec.~\ref{sec:cigar} outlines the CIGAR analysis; Sec.~\ref{sec:analysis} presents the results; and Sec.~\ref{sec:summary} summarizes our 
conclusions.

\section{The JAM model and simulations at $\sqrt{s_{\rm NN}} = 2.4$~GeV}\label{sec:jam_2p4}

The Jet AA Microscopic Transport Model (JAM) is a nonequilibrium hadronic transport approach 
applicable over a broad range of collision energies~\cite{2019JAM,PhysRevC.61.024901}. 
It propagates hadrons and resonances through successive microscopic interactions and includes 
energy-dependent mechanisms for inelastic hadron-hadron collisions, such as resonance 
excitation at low energies and string excitation at higher energies. JAM has previously been 
employed to investigate proton-number fluctuations and the influence of hadronic processes on 
higher-order cumulants~\cite{He2016_JAM,PhysRevC.101.034909,Chatterjee_2021}.

We simulate Au+Au (\(A=197\)) and Ag+Ag (\(A=107\)) collisions at 
\(\sqrt{s_{\mathrm{NN}}}=2.4\ \mathrm{GeV}\) using both the cascade and mean-field versions of 
JAM~\cite{hades_2p4}. The impact parameter is sampled over \(0\le b\le 16\ \mathrm{fm}\). 
In cascade mode (\texttt{MeanField:mode = 0}), hadrons propagate between binary scatterings 
and resonance decays without an explicit mean-field potential. In mean-field mode, their 
propagation is additionally modified by a density-dependent nuclear potential. The 
single-particle Hamiltonian and the corresponding equations of motion are given in Eqs.~\eqref{eq:Hamiltonian} 
and~\eqref{eq:HamiltonianEquation} of the original manuscript. The interaction potential, 
Eq.~\eqref{eq:PotentialEnergy}, contains isoscalar density-dependent and isovector symmetry terms.

\begin{equation}
H_i = \sqrt{\vec{p}_i^2 + m_{\mathrm{eff}}^2} + U(\vec{r}_i),
\label{eq:Hamiltonian}
\end{equation}
where \(\vec{p}_i\) is the momentum, \(m_{\mathrm{eff}}\) denotes the mass of stable hadrons 
and, for resonances, the effective mass determined by the resonance mass distribution, and 
\(U(\vec{r}_i)\) is a coordinate-dependent potential.
\begin{equation}
\frac{d\vec{r}_i}{dt} = \frac{\partial H_i}{\partial \vec{p}_i}
= \frac{\vec{p}_i}{\sqrt{\vec{p}_i^2 + m_{\mathrm{eff}}^2}}, \qquad
\frac{d\vec{p}_i}{dt} = -\frac{\partial H_i}{\partial \vec{r}_i}
= -\frac{\partial U}{\partial \vec{r}_i}
\label{eq:HamiltonianEquation}
\end{equation}
The mean-field potential is calculated as a function of the local density,
\begin{equation}
U(\rho,\rho_{I_3}) = a\left(\frac{\rho}{\rho_0}\right)
+ b\left(\frac{\rho}{\rho_0}\right)^\tau
\pm 2S_{\mathrm{pot}}\frac{\rho_{I_3}}{\rho_0}
\label{eq:PotentialEnergy}
\end{equation}

In Eq.~\eqref{eq:PotentialEnergy}, \(\rho\) is the local baryon density evaluated in the Eckart 
frame, \(\rho_{I_3}\) is the corresponding isospin density, and \(\rho_0=0.168\ \mathrm{fm}^{-3}\) 
is the saturation density. The symmetry-potential contribution has opposite signs for 
neutrons and protons. The potential acts on baryons; mesonic mean fields and electromagnetic 
Coulomb and Lorentz forces are not included in this setup. We use the soft, 
momentum-independent nuclear equation of state with incompressibility \(K=210\ \mathrm{MeV}\) 
and parameters \(a=-356\ \mathrm{MeV}\), \(b=303\ \mathrm{MeV}\), \(\tau=7/6\), and 
\(S_{\mathrm{pot}}=18\ \mathrm{MeV}\)~\cite{PhysRevC.105.014911,
PhysRevC.72.064908}. These settings follow Ref.~\cite{like_25}. 

\begin{figure}[htbp]
    \centering
    \includegraphics[scale=0.5]{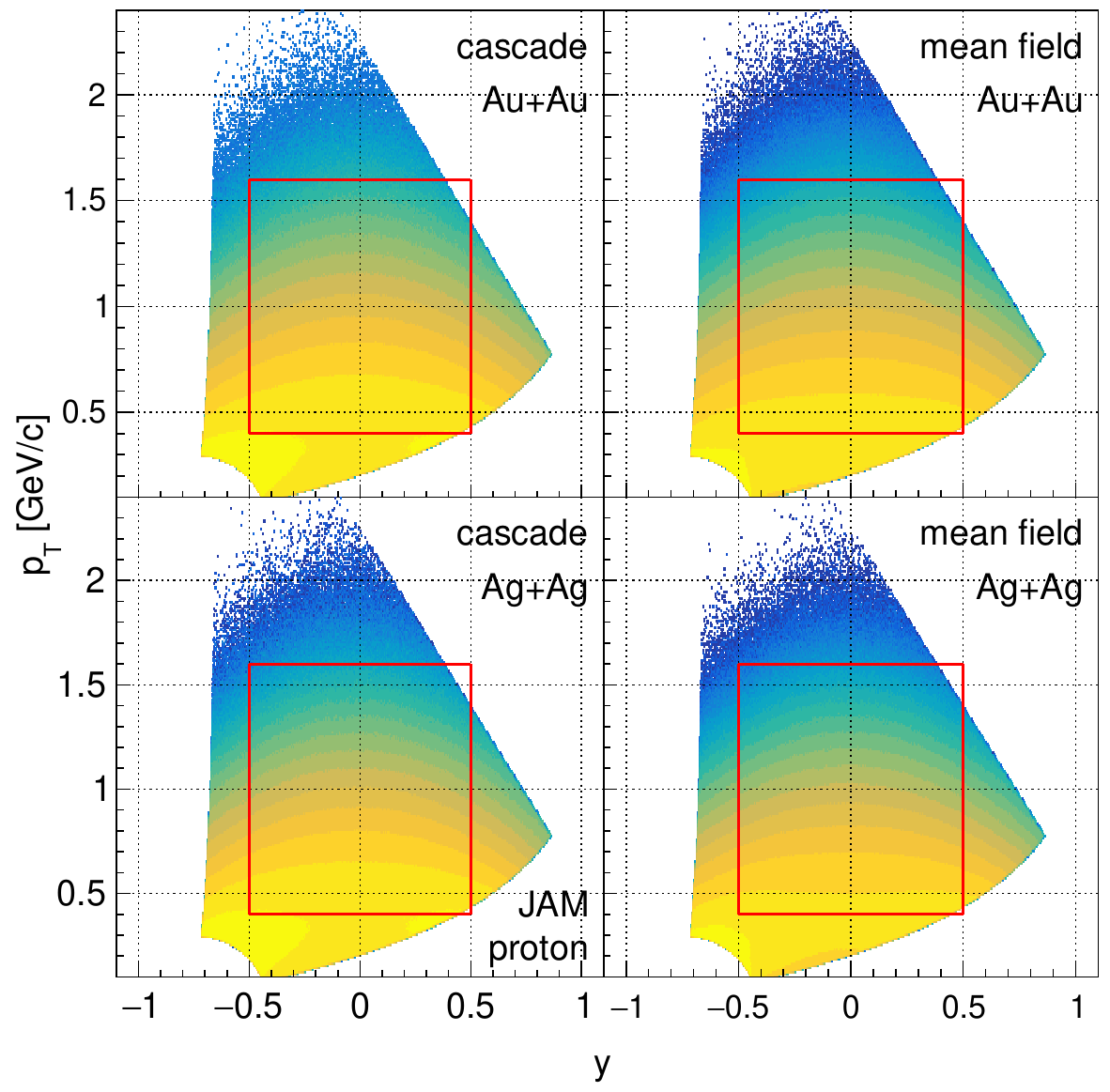}
    \caption{
        Proton transverse momentum $p_T$ versus center-of-mass rapidity
        $y$ in Au+Au and Ag+Ag collisions at
        $\sqrt{s_{\rm NN}}=2.4$~GeV, calculated with the JAM model. 
        The upper (lower) panels show Au+Au (Ag+Ag), and the left (right) panels show cascade 
        (mean-field) mode. Red rectangles mark \(0.4<p_T<1.6\ \mathrm{GeV}/c\) and \(-0.5<y<0.5\) 
        corresponding to \(y_{\mathrm{lab}}=0.74\pm0.5\).
    }
    \label{fig:PtY}
\end{figure}

The HADES Collaboration has measured proton-number fluctuations in Au+Au
collisions at \(\sqrt{s_{\mathrm{NN}}}=2.4\,\mathrm{GeV}\) at
GSI~\cite{hades_2p4}. To compare our JAM calculations with these data, we
adopt the same proton acceptance as in the HADES analysis. In the
fixed-target laboratory frame, protons are selected with
\(15^\circ<\theta_{\mathrm{lab}}<85^\circ\) and
\(0.3<p<3\,\mathrm{GeV}/c\), and the analysis of proton-number fluctuations is performed in
\(0.4<p_T<1.6\,\mathrm{GeV}/c\) and
\(y_{\mathrm{lab}}=y_0\pm\Delta y\), with \(y_0=0.74\) and
\(\Delta y=0.1\)--\(0.5\). Here and in what follows, \(y\) denotes the
rapidity in the nucleon-nucleon center-of-mass frame,
\(y\equiv y_{\mathrm{cm}}=y_{\mathrm{lab}}-y_0\), where
\(y_{\mathrm{lab}}\) is the laboratory-frame rapidity and \(y_0=0.74\)
is the rapidity of the nucleon-nucleon center-of-mass frame in the
laboratory; the interval \(y_{\mathrm{lab}}=y_0\pm\Delta y\) therefore
corresponds to \(-\Delta y<y<\Delta y\).
Figure~\ref{fig:PtY} shows the accepted phase space in the
\(p_T\)--\(y\) plane for Au+Au and Ag+Ag collisions at
\(\sqrt{s_{\mathrm{NN}}}=2.4\,\mathrm{GeV}\) in both cascade and
mean-field modes. The red boxes indicate the largest rapidity window used
in the present analysis; varying \(\Delta y\) within this window allows us
to study the rapidity dependence of the proton cumulants. The angular cut
corresponds to \(0.087<\eta_{\mathrm{lab}}<2.028\). Since the HADES
results and their physics implications were discussed in
Sec.~\ref{sec:introduction}, here we focus on the JAM implementation and
the acceptance used for the comparison.

\begin{table}[]
\begin{tabular}{@{}ccccc@{}}
\toprule
mode            & cascade & mean-field \\ \midrule
Events(Au+Au) /million    & 50  & 23  \\
Events(Ag+Ag) / million & 36  & 23  \\ \bottomrule
\end{tabular}
\caption{Numbers of generated JAM events (in millions) for Au+Au and Ag+Ag collisions at \(\sqrt{s_{\mathrm{NN}}}=2.4\ \mathrm{GeV}\) in cascade and mean-field modes.} 
\label{tab:tabel_events}
\end{table}

The event samples comprise 50 million Au+Au and 36 million Ag+Ag collisions in cascade mode, 
and 23 million events for each system in mean-field mode (Table~\ref{tab:tabel_events}). Unless stated otherwise, 
the fluctuation analysis uses the acceptance \(0.4<p_T<1.6\ \mathrm{GeV}/c\) and \(-0.5<y<0.5\).

\section{Moments, cumulants, and factorial cumulants}
\label{sec:moments}

The shape of a probability distribution \(P(N)\) can be characterized by its moments. 
The variance \(\sigma^2\) measures the width of the distribution, while the skewness \(S\) 
and kurtosis \(Kurt\) quantify, respectively, its asymmetry and its peakedness around the mean.

For an event-by-event proton multiplicity \(N\) with probability distribution \(P(N)\), 
the moment-generating function is defined as
\begin{equation}
M(t)=\langle e^{tN}\rangle .
\end{equation}
The \(n^{th}\)-order moment \(\langle N^n\rangle\) is obtained from the Taylor expansion of 
\(M(t)\) around \(t=0\):
\begin{equation}
M(t)=\langle e^{tN}\rangle
=1+t\langle N\rangle+\frac{t^2}{2!}\langle N^2\rangle
+\frac{t^3}{3!}\langle N^3\rangle+\cdots+\frac{t^n}{n!}\langle N^n\rangle .
\end{equation}

Introducing the deviation from the mean, \(\delta N=N-\langle N\rangle\), 
the \(n\)th-order central moment is defined as
\begin{equation}
\mu_n=\langle(\delta N)^n\rangle,\qquad n\ge 2 .
\end{equation}

The cumulant generating function is
\begin{equation}
G(t)=\ln M(t)=\ln\langle e^{tN}\rangle,
\end{equation}
and the \(n\)th-order cumulant \(C_n\) is defined by the Taylor expansion
\begin{equation}
G(t)=\sum_{n=1}^{\infty} C_n\frac{t^n}{n!}
=C_1 t+C_2\frac{t^2}{2!}+\cdots .
\label{eq:cumulant-expansion}
\end{equation}

The first four cumulants are related to the central moments by
\begin{equation}
\label{eq:cumulants}
\begin{split}
C_1&=\langle N\rangle,\\
C_2&=\langle(\delta N)^2\rangle=\mu_2,\\
C_3&=\langle(\delta N)^3\rangle=\mu_3,\\
C_4&=\langle(\delta N)^4\rangle-3\langle(\delta N)^2\rangle^2
=\mu_4-3\mu_2^2 .
\end{split}
\end{equation}
For \(n>3\), the cumulants can be obtained recursively:
\begin{equation}
C_n=\mu_n-\sum_{m=2}^{n-2}\binom{n-1}{m-1}C_m\mu_{n-m} .
\end{equation}

The mean, variance, skewness, and excess kurtosis are then
\begin{equation}
\label{eq:standardized}
\begin{split}
Mean&=\langle N\rangle=C_1,\\
\sigma^2&=\langle(\delta N)^2\rangle=C_2,\\
S&=\frac{\langle(\delta N)^3\rangle}{\sigma^3}
=\frac{C_3}{C_2^{3/2}},\\
Kurt&=\frac{\langle(\delta N)^4\rangle}{\sigma^4}-3
=\frac{C_4}{C_2^2} .
\end{split}
\end{equation}

The first two cumulants are the mean and variance, \(C_1=\langle N\rangle\) and \(C_2=\sigma^2\). 
The third and fourth cumulants quantify the asymmetry and excess kurtosis of the distribution, 
through \(S=C_3/C_2^{3/2}\) and \(\mathrm{Kurt}=C_4/C_2^2\) in Eqs~\eqref{eq:standardized}.

Ratios of cumulants are useful because the leading extensive contribution can cancel under 
suitable assumptions. The commonly studied ratios are
\begin{equation}
\label{eq:cumulant-ratios}
\frac{C_2}{C_1}=\frac{\sigma^2}{Mean},
\qquad
\frac{C_3}{C_2}=S\sigma,
\qquad
\frac{C_4}{C_2}=Kurt\,\sigma^2 .
\end{equation}
Here, we use \(C_3/C_1\) in place of \(C_3/C_2\), which can be written as 
\(C_3/C_1=S\sigma^3/Mean\), to facilitate a common normalization by the 
mean proton number. We report \(C_2/C_1\), \(C_3/C_1\), and \(C_4/C_2\) throughout.

Factorial cumulants \(K_n\) related to \(C_n\) by 
\begin{equation}
\label{eq:factorial-cumulants}
\begin{split}
\kappa_1&=C_1=\langle N\rangle,\\
\kappa_2&=-C_1+C_2,\\
\kappa_3&=2C_1-3C_2+C_3,\\
\kappa_4&=-6C_1+11C_2-6C_3+C_4,
\end{split}
\end{equation}
quantify connected multiplicity correlations of a given order.
For a Poisson distribution, \(\kappa_n=0\) for all \(n\ge2\). 
Thus, nonzero higher-order factorial cumulants provide a direct measure of deviations 
from an independent-emission Poisson baseline. We report the normalized quantities 
\(\kappa_2/\kappa_1\), \(\kappa_3/\kappa_1\), and \(\kappa_4/\kappa_1\) alongside the ordinary 
cumulant ratios.They therefore isolate genuine \(n\)-particle correlations beyond the Poisson 
baseline, which is particularly useful when analyzing proton-number fluctuations in 
heavy-ion collisions.

\section{Centrality-Independent Genuine Cumulant Analysis Framework (CIGAR)}
\label{sec:cigar}

We employ the centrality-independent analysis
framework of Ref.~\cite{WANG2025139984}, referred to as CIGAR in the present
study, to mitigate biases associated with event-by-event fluctuations of the 
initial collision volume. The effects of initial-volume fluctuations are more
pronounced in low-energy heavy-ion collisions than at high energies. Rather than assigning each event to a centrality class using a 
single measured multiplicity estimator, the framework models the observed proton-number 
distribution as a superposition of distributions associated with narrower centrality intervals. 
The parameters of these component distributions are inferred from the full multiplicity 
distribution.

Statistically, the probability density function \(p(x)\) of a distribution can
be approximated by a Gram--Charlier series expressed in terms of its cumulants.
As a practical solution to the moment problem, the Edgeworth expansion
approximates \(p(x)\) around a Gaussian distribution and incorporates
higher-order cumulants to refine the shape of the distribution.

Let \(\phi(x)\) denote the standard normal density,
\begin{equation}
\phi(x) = \frac{1}{\sqrt{2\pi}} e^{-x^2/2}.
\end{equation}
The Gram--Charlier series can be written as
\begin{equation}
p(x) = \sum_{n=0}^{\infty} c_n \frac{d^n \phi(x)}{dx^n},
\label{eq:GaussianExpanding}
\end{equation}
where the coefficients \(c_n\) are given by
\begin{equation}
c_n = \frac{(-1)^n}{n!} \int_{-\infty}^{\infty} p(x) \, \mathrm{He}_n(x) \, dx.
\label{eq:FourierTransform}
\end{equation}
The Chebyshev--Hermite polynomials \(\mathrm{He}_n(x)\) are defined as
\begin{equation}
\mathrm{He}_n(x) = (-1)^n e^{x^2/2} \frac{d^n}{dx^n} e^{-x^2/2}.
\label{eq:ChebyshevHermitePolynomials}
\end{equation}

Equivalently, the Edgeworth expansion expresses \(p(x)\) directly in terms of
the standardized cumulants \(\lambda_n = C_n / \sigma^n\), where \(C_n\) is the
\(n\)th cumulant in the notation of this paper and \(\sigma\) is the standard
deviation. Up to fourth order, the expansion reads
\begin{equation}
p(x) = \frac{1}{\sigma} \phi\!\left(\frac{x-\mu}{\sigma}\right)
\left[ 1 + \frac{\lambda_3}{6} \mathrm{He}_3\!\left(\frac{x-\mu}{\sigma}\right)
+ \frac{\lambda_4}{24} \mathrm{He}_4\!\left(\frac{x-\mu}{\sigma}\right)
+ \frac{\lambda_3^2}{72} \mathrm{He}_6\!\left(\frac{x-\mu}{\sigma}\right) + \cdots \right],
\label{eq:EdgeworthExpansion}
\end{equation}
where \(\mu = \langle N\rangle\) is the mean, \(\lambda_3 = C_3/\sigma^3\) is
the skewness, and \(\lambda_4 = C_4/\sigma^4\) is the excess kurtosis.

To reconstruct the proton-number distribution for a given centrality interval,
CIGAR optimizes a set of parameters such that the sum of all sub-distributions
closely matches the simulated or measured total distribution. A differential
evolution algorithm is employed for the global search of the parameter space. Within this framework, the cumulants and their ratios are
parameterized as functions of the centrality bin index \(n\) (ordered from the
most central to the most peripheral interval):
\begin{equation}
\begin{split}
    C_1 &= a_{C_1} n^4 + b_{C_1} n^3 + c_{C_1} n^2 + d_{C_1} n + e_{C_1},\\
    C_2 &= a_{C_2} + b_{C_2} C_1 + c_{C_2} C_1^2,\\
    C_3 &= a_{C_3} + b_{C_3} C_1 + c_{C_3} C_1^2,\\
    C_4/C_2 &= a_{C_4/C_2} + b_{C_4/C_2} C_1 + c_{C_4/C_2} C_1^2 .
\end{split}
\label{eq:cumPara}
\end{equation}
Equation~\eqref{eq:cumPara} shows the adopted parameterization strategy, in
which the cumulants are described by low-order polynomials. This approach
reduces the number of free parameters, stabilizes the optimization, and
preserves the physical interpretation of the extracted cumulants.

\begin{figure}[t]
    \centering
    \includegraphics[scale=0.6]{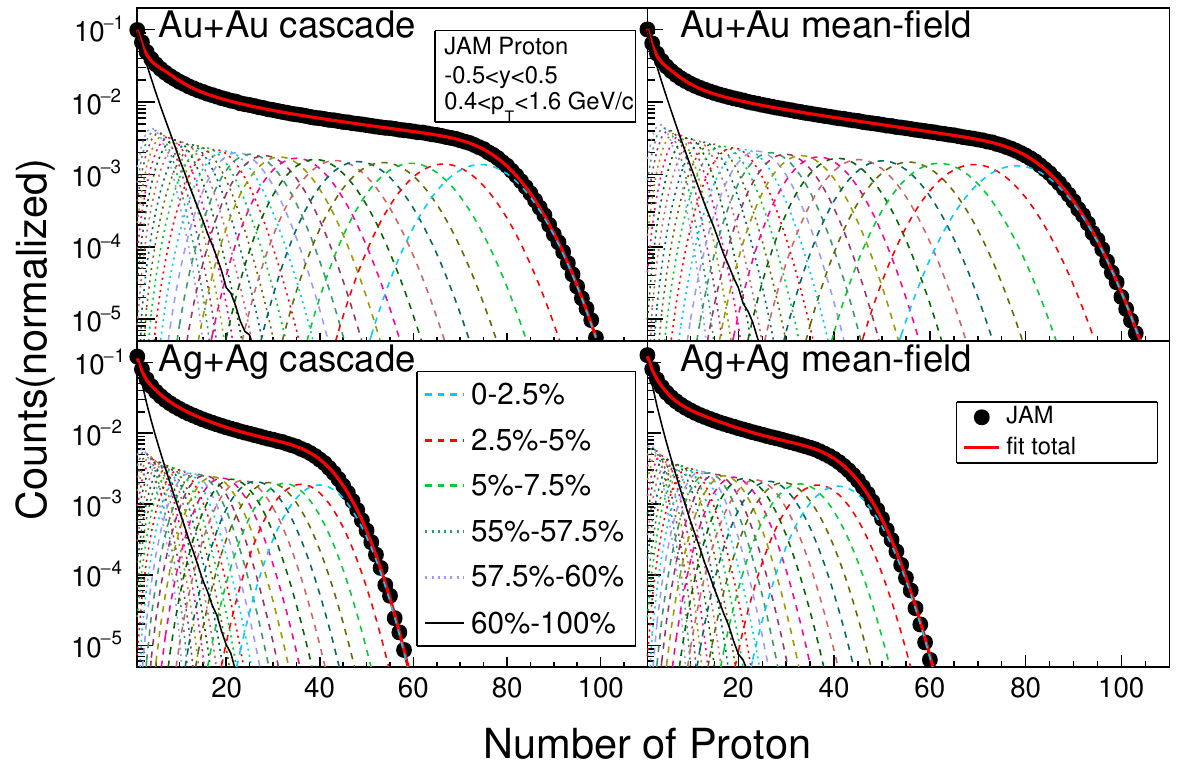}
    \caption{
        Proton-number distributions and fit results in Au+Au and Ag+Ag
        collisions at \(\sqrt{s_{\rm NN}} = 2.4\)~GeV from the JAM model.
        The first and second rows correspond to Au+Au and Ag+Ag collisions,
        respectively, and the first and second columns correspond to cascade
        and mean-field modes, respectively. Black solid circles represent the
        proton-number distributions from JAM. Colored dashed lines show the
        fitted sub-distributions for consecutive \(2.5\%\) centrality bins,
        from \(0\)--\(2.5\%\) to \(57.5\)--\(60\%\). The black solid line
        indicates the \(60\%\)--\(100\%\) peripheral events, and the red solid
        line is the sum of all sub-distributions. The proton transverse
        momentum and rapidity windows are \(0.4 < p_T < 1.6\)~GeV/\(c\) and
        \(-0.5 < y < 0.5\), respectively.
    }
    \label{fig:pdfNp}
\end{figure}

Figure~\ref{fig:pdfNp} compares the input JAM proton-number distributions with their 
reconstructed counterparts for both collision systems and both JAM modes.
within the acceptance \(0.4<p_T<1.6\)~GeV/\(c\) and \(-0.5<y<0.5\). The
black solid circles represent the proton-number distributions from the JAM
model. The colored dashed lines each correspond to a fitted sub-distribution
over a \(2.5\%\) centrality interval, ranging from \(0\)--\(2.5\%\) to
\(57.5\)--\(60\%\). The black solid line represents the \(60\%\)--\(100\%\)
peripheral events, and the red solid line is the sum of all individual
contributions, i.e., the total fitted distribution. From the fitted parameters
of these sub-distributions, the cumulants are extracted for each centrality
interval and are presented in Sec.~\ref{sec:analysis}.

In addition to CIGAR method, we also performed calculation with the conventional method using centrality bin width correction (CBWC)~\cite{Luo:2013bmi}. The CBWC calculation are performed with both $N_{\rm part}$ and RefMult3 which is a reference multiplicity excluding protons to avoid auto-correlation effect~\cite{Luo:2013bmi}. The $N_{\rm part}$-based CBWC results are used as baseline calculations without volume fluctuations.

\section{Results and Discussion}\label{sec:analysis}

\begin{figure}[htbp]
    \centering
    \includegraphics[scale=0.8]{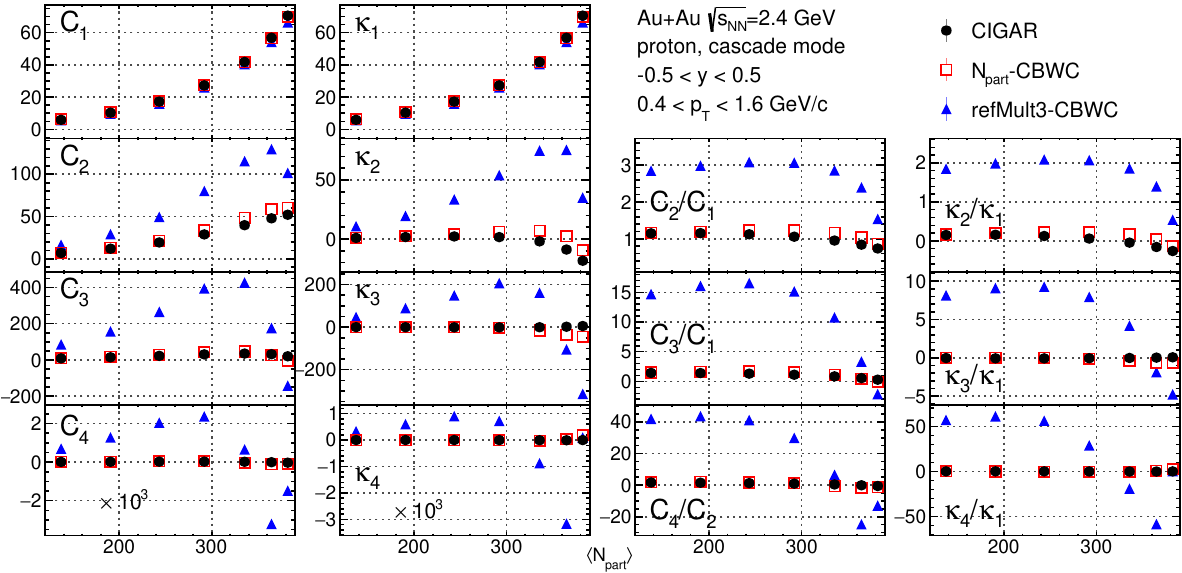}
    \caption{
        Centrality dependence of the proton-number cumulants ($C_{1}$, $C_{2}$,
        $C_{3}$, and $C_{4}$), factorial cumulants ($\kappa_{1}$,
        $\kappa_{2}$, $\kappa_{3}$, and $\kappa_{4}$), cumulant ratios
        ($C_{2}/C_{1}$, $C_{3}/C_{1}$, and $C_{4}/C_{2}$), and factorial
        cumulant ratios ($\kappa_{2}/\kappa_{1}$,
        $\kappa_{3}/\kappa_{1}$, and $\kappa_{4}/\kappa_{1}$) from the JAM
        model in cascade mode for Au+Au collisions at
        $\sqrt{s_{\rm NN}}=2.4$~GeV. 
        Results from CIGAR (Black solid circles) are compared with CBWC using $N_{\rm part}$
        (red open squares) and RefMult3 (blue solid triangles). 
        The acceptance is \(-0.5<y<0.5\) and \(0.4<p_{T}<1.6\)~GeV/\(c\).
    }
    \label{fig:cumKap_cent_ym0p5p0p5}
\end{figure}

Figure~\ref{fig:cumKap_cent_ym0p5p0p5} compares the centrality dependence of the
$C_{1}$--$C_{4}$, $\kappa_{1}$--$\kappa_{4}$, the ordinary cumulant ratios, and the corresponding 
factorial cumulant ratios for Au+Au collisions at
$\sqrt{s_{\rm NN}}=2.4$~GeV from the JAM model in cascade mode, within
\(-0.5<y<0.5\) and \(0.4<p_{T}<1.6\)~GeV/\(c\). The $N_{\rm part}$-based CBWC
results (red open squares) are used as a benchmark, since centrality
bins defined by $N_{\rm part}$ are least affected by fluctuations of the
centrality estimator. The RefMult3-based CBWC results (blue solid triangles)
and the CIGAR results (black solid circles) are then compared with this
benchmark. In all panels and for all centrality bins, the CIGAR results closely
follow the $N_{\rm part}$-based benchmark, including for the higher-order observables. 
In contrast, the RefMult3-based CBWC results systematically deviate from the $N_{\rm part}$
benchmark. At this low collision energy, charged-pion and charged-kaon yields are relatively 
small, reducing the multiplicity available for centrality determination and degrading the 
resolution of a reference-multiplicity estimator. Residual fluctuations in the number of 
participants can therefore significantly affect the conventionally corrected cumulants.
The agreement between CIGAR and the \(N_{\mathrm{part}}\)-based benchmark supports the use of 
the reconstruction method to reduce such biases arising from initial-volume fluctuations in 
low-energy proton-number fluctuation analyses.

This comparison provides a useful closure test within the chosen JAM sample. 
It provides qualitative support that the CIGAR framework remains effective in the low-energy, 
low-multiplicity regime at $\sqrt{s_{\rm NN}}=2.4$~GeV, where initial-volume fluctuations are most pronounced, 
and indicates that the method is applicable to experimental multiplicity distributions in the 
energy region accessed by HADES.
However, quantifying the residual systematic uncertainty of the reconstructed cumulants is beyond 
the scope of the present work and would additionally require tests of the Edgeworth truncation, 
parameterization flexibility, reconstruction stability, and possible model dependence. Such 
checks are particularly relevant before applying the method to experimental multiplicity 
distributions in future analyses.

\begin{figure}[htbp]
    \centering
    \includegraphics[scale=0.8]{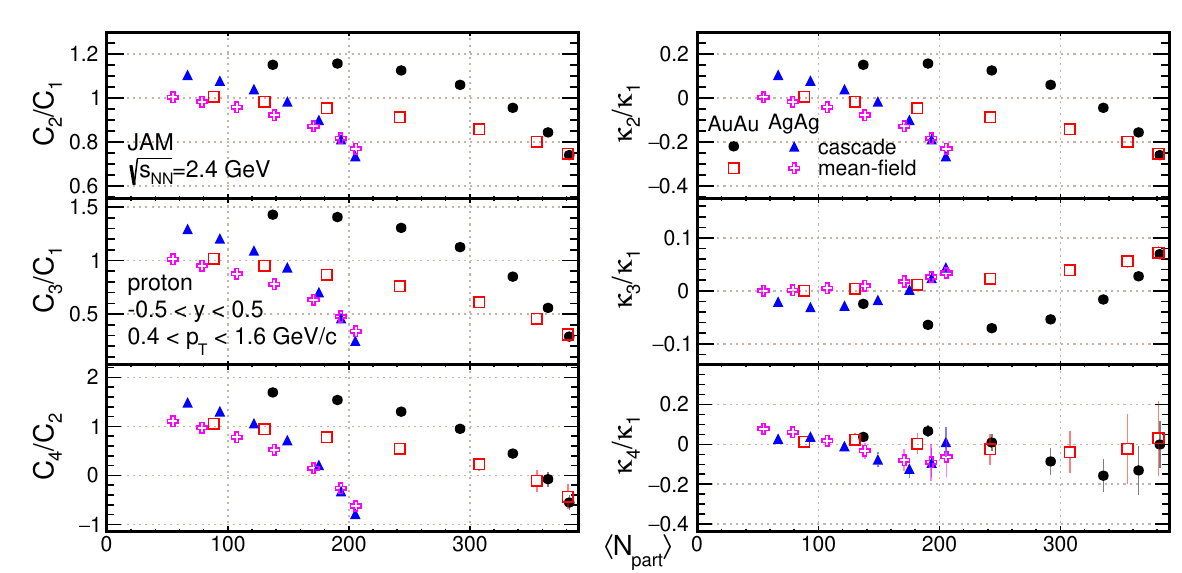}
    \caption{
        Proton-number cumulant ratios ($C_{2}/C_{1}$, $C_{3}/C_{1}$, $C_{4}/C_{2}$)
        and factorial cumulant ratios ($\kappa_{2}/\kappa_{1}$,
        $\kappa_{3}/\kappa_{1}$, $\kappa_{4}/\kappa_{1}$) from the JAM model
        at $\sqrt{s_{\rm NN}}=2.4$~GeV as a function of
        $\langle N_{\rm part}\rangle$. Symbols represent Au+Au cascade (black solid circles), 
        Au+Au mean field (red open squares), Ag+Ag cascade (blue sold triangles), and Ag+Ag 
        mean field (magenta open crosses). The acceptance is \(-0.5<y<0.5\) and
        \(0.4<p_{T}<1.6\)~GeV/\(c\).
    }
    \label{fig:cumKap_cent_sysModel_ym0p5p0p5}
\end{figure}

Figure~\ref{fig:cumKap_cent_sysModel_ym0p5p0p5} shows the centrality
dependence of the cumulant ratios ($C_{2}/C_{1}$, $C_{3}/C_{1}$, and
$C_{4}/C_{2}$) and factorial cumulant ratios ($\kappa_{2}/\kappa_{1}$,
$\kappa_{3}/\kappa_{1}$, and $\kappa_{4}/\kappa_{1}$) as a function of 
$\langle N_{\rm part}\rangle$ from the JAM model at $\sqrt{s_{\rm NN}}=2.4$~GeV,
within \(-0.5<y<0.5\) and \(0.4<p_{T}<1.6\)~GeV/\(c\). Black solid circles,
red open boxes, blue solid triangles, and magenta open crosses correspond to
Au+Au cascade, Au+Au mean-field, Ag+Ag cascade, and Ag+Ag mean-field,
respectively. For a given centrality interval, the cascade-mode results extend
to larger $\langle N_{\rm part}\rangle$ than the mean-field results, and this
difference is more pronounced in Au+Au than in Ag+Ag collisions. The
$\langle N_{\rm part}\rangle$ values in Au+Au collisions are systematically
larger than those in Ag+Ag collisions, as expected from the larger mass number
of Au compared with Ag. For all four simulation configurations,
$C_{2}/C_{1}$, $C_{3}/C_{1}$, and $C_{4}/C_{2}$ generally decrease as the collisions 
become more central. 

The nuclear mean field produces observable changes in several cumulant ratios.
In the 10--60\% centrality region, $C_{2}/C_{1}$, $C_{3}/C_{1}$, $C_{4}/C_{2}$,
and $\kappa_{2}/\kappa_{1}$ are larger in cascade mode than in the mean-field. 
In the most central 0--5\% events, by contrast, the mean-field values of $C_{2}/C_{1}$,
$C_{3}/C_{1}$, and $\kappa_{2}/\kappa_{1}$ exceed the corresponding cascade values, whereas
the mean-field values of $C_{4}/C_{2}$ and $\kappa_{3}/\kappa_{1}$ are lower. Within the 
statistical uncertainties, $\kappa_{4}/\kappa_{1}$ shows no robust separation between the two transport modes. 
Similar decreases of higher-order cumulant ratios toward central events have been 
observed in other hadronic-transport calculations~\cite{Zhang_2026,xu2026protonhighordercumulantsauau}.
Such behavior can arise from the interplay of baryon-number conservation, 
the fraction of baryons contained within the measured acceptance, and the hadronic collision dynamics.

In peripheral collisions, the mean-field values of $C_{2}/C_{1}$, $C_{3}/C_{1}$, and
$C_{4}/C_{2}$ are consistent with unity for both collision systems and are
lower than the corresponding cascade values. The behavior of
$\kappa_{2}/\kappa_{1}$ is similar to that of $C_{2}/C_{1}$: its peripheral
mean-field values are close to zero and lower than the cascade values.
The differences between the cascade and mean-field results can be understood
as follows. In cascade mode, nucleons propagate along straight-line
trajectories between collisions and the nuclear mean-field potential is
absent. In mean-field mode, the attractive nuclear potential and
momentum-independent interactions modify the expansion dynamics, leading to a
more pronounced collective flow and a different freeze-out
geometry~\cite{PhysRevC.105.014911,PhysRevC.101.034909}. A recent study has
shown that at 2.4~GeV the fraction of protons preserving their initial isospin
is higher in mean-field mode, whereas the fraction originating from isospin
conversion is higher in cascade mode~\cite{fp85-w4l2}. These different
proton-source compositions partly account for the observed mode dependence of
the cumulant ratios.

At \(\sqrt{s_{\mathrm{NN}}}=2.4\ \mathrm{GeV}\), a phenomenological freeze-out 
parameterization gives \(\mu_B=790\ \mathrm{MeV}\)~\cite{PhysRevC.73.034905}, placing the system in 
a region where nuclear-matter correlations and possible nuclear liquid-gas transition 
effects may become relevant. Calculations based on other theoretical approaches have found 
that nucleon interactions and liquid-gas dynamics can affect higher-order baryon-number 
cumulants in this energy domain~\cite{recchi2026imprintsnuclearliquidgasphase}. 
The present JAM results establish sensitivity to hadronic interactions but do not by 
themselves identify a phase transition; a dedicated comparison between transport dynamics and 
models with controlled phase-transition physics would be required.

\begin{figure}[htbp]
    \centering
    \includegraphics[scale=0.8]{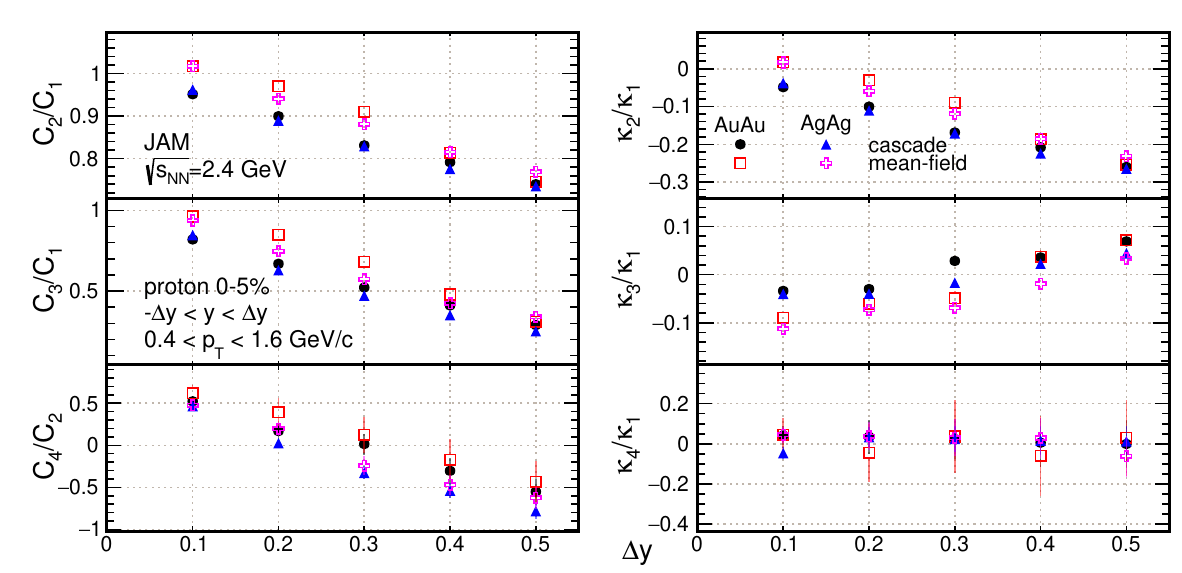}
    \caption{
        Rapidity dependence of the proton-number cumulant ratios
        ($C_{2}/C_{1}$, $C_{3}/C_{1}$, $C_{4}/C_{2}$) and factorial cumulant
        ratios ($\kappa_{2}/\kappa_{1}$, $\kappa_{3}/\kappa_{1}$,
        $\kappa_{4}/\kappa_{1}$) from the JAM model at
        $\sqrt{s_{\rm NN}}=2.4$~GeV. Symbol definitions are the same as in Figure~\ref{fig:cumKap_cent_sysModel_ym0p5p0p5}.
        The transverse-momentum range is \(0.4<p_{T}<1.6\)~GeV/\(c\), and the
        rapidity window is \(-\Delta y<y<\Delta y\) with $\Delta y$ ranging
        from \(0.1\) to \(0.5\).
    }
    \label{fig:cumKap_rap}
\end{figure}

Figure~\ref{fig:cumKap_rap} presents the rapidity dependence of the cumulant
ratios ($C_{2}/C_{1}$, $C_{3}/C_{1}$, and $C_{4}/C_{2}$) and factorial
cumulant ratios ($\kappa_{2}/\kappa_{1}$, $\kappa_{3}/\kappa_{1}$, and
$\kappa_{4}/\kappa_{1}$) from the JAM model at
$\sqrt{s_{\rm NN}}=2.4$~GeV, for \(0.4<p_{T}<1.6\)~GeV/\(c\) and
\(-\Delta y<y<\Delta y\) with $\Delta y=0.1$--\(0.5\). Black solid circles,
red open boxes, blue solid triangles, and magenta open crosses indicate
Au+Au cascade, Au+Au mean-field, Ag+Ag cascade, and Ag+Ag mean-field,
respectively. For $C_{2}/C_{1}$, $C_{3}/C_{1}$, $C_{4}/C_{2}$, and
$\kappa_{2}/\kappa_{1}$, all four simulation configurationss exhibit a decreasing trend with
increasing $\Delta y$. For both Au+Au and Ag+Ag collisions, these four ratios in
mean-field mode are larger than the corresponding cascade-mode values, except
in the range $-0.1<y<0.1$, where $C_{4}/C_{2}$ in cascade mode is essentially
the same as that in mean-field mode. In contrast, $\kappa_{3}/\kappa_{1}$
increases with $\Delta y$ for all four simulation configurations, and its mean-field values
are lower than the cascade-mode values in both collision systems. This contrasting behavior may partly reflect the different sensitivities of
the two-particle and three-particle cumulants to the mean-field-driven
collective expansion and to the proton-source composition, which differs
between the two modes~\cite{PhysRevC.105.014911,fp85-w4l2}; a quantitative
attribution, however, requires further study. For $C_{3}/C_{1}$ and $C_{4}/C_{2}$, the Ag+Ag values are slightly lower than the Au+Au values, although the
difference is small and, for the largest rapidity window, comparable to the
statistical uncertainty;
a weaker trend is seen for $\kappa_{3}/\kappa_{1}$. For $\kappa_{4}/\kappa_{1}$, no clear rapidity dependence can be established, and
all values are consistent with zero within statistical uncertainties. 

These system-dependent differences may partly arise from the smaller number of participating nucleons in Ag+Ag
collisions, which can reduce the baryon-number-conservation constraint and
modify the freeze-out dynamics, and may also reflect the different effective
rapidity coverage of multi-particle correlations. A quantitative attribution requires further study. These differences are, however, small and not significant at the current
statistical accuracy; in the largest rapidity window the two systems agree
within about one standard deviation. We therefore do not assign strong significance to these differences.
As the rapidity acceptance grows, a larger fraction of the baryons participates in the 
measured proton multiplicity, increasing sensitivity to global conservation constraints and to 
correlations spanning the accepted phase space~\cite{Zhang_2026,fp85-w4l2}. Changes in proton-source composition 
and freeze-out dynamics may further influence the observed acceptance dependence.The distinct 
trends of $\kappa_{2}/\kappa_{1}$ and $\kappa_{3}/\kappa_{1}$ suggest that different correlation orders respond 
differently to these mechanisms, but a quantitative decomposition would require additional 
differential studies.

\begin{figure}[htbp]
    \centering
    \includegraphics[scale=0.8]{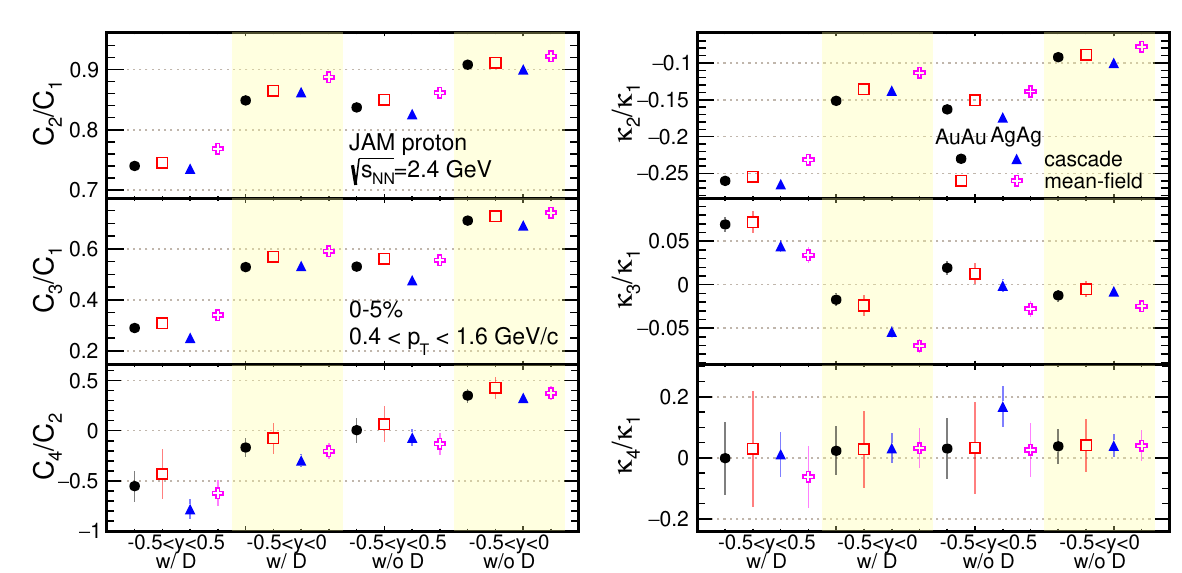}
    \caption{
        Proton-number cumulant ratios ($C_{2}/C_{1}$, $C_{3}/C_{1}$, and $C_{4}/C_{2}$) and
        factorial cumulant ratios ($\kappa_{2}/\kappa_{1}$,
        $\kappa_{3}/\kappa_{1}$, and $\kappa_{4}/\kappa_{1}$) in central
        collisions from the JAM model at $\sqrt{s_{\rm NN}}=2.4$~GeV, for
        \(0.4<p_{T}<1.6\)~GeV/\(c\). Black solid circles, red open boxes, blue
        solid triangles, and magenta open crosses indicate Au+Au cascade,
        Au+Au mean-field, Ag+Ag cascade, and Ag+Ag mean-field, respectively.
        The abscissa is divided into four groups: (1) $-0.5<y<0.5$, w/ D;
        (2) $-0.5<y<0$, w/ D; (3) $-0.5<y<0.5$, w/o D; and (4) $-0.5<y<0$,
        w/o D. Here ``w/ D'' denotes the original JAM results without
        light-nuclei formation, and ``w/o D'' denotes the results after
        applying an efficiency factor to mimic the removal of free protons
        due to deuteron-like light-nuclei formation.
    }
    \label{fig:cumKap_sysModeHalfDeu}
\end{figure}

Light nuclei are not produced in the present JAM simulations. Previous work
has shown that the effect of light nuclei such as deuterons on proton cumulants
is similar to that of an efficiency~\cite{Chatterjee_2021}. To estimate the 
possible depletion of measured free protons by cluster formation, we perform a 
simple test in which a fraction of the protons is removed
event by event according to an efficiency factor. For Au+Au collisions at
$\sqrt{s_{\rm NN}}=2.4$~GeV, the yields of protons, deuterons, tritons, and
other particles measured by HADES are listed in Table~I of
Ref.~\cite{MOTORNENKO2021136703}. Since deuterons are the most abundant species
after protons, we estimate the fraction of protons remaining as free protons
after deuteron formation from the measured deuteron-to-proton ratio,
$\epsilon=1-d/p$, obtaining $\epsilon=0.63$ for Au+Au collisions. For Ag+Ag
collisions at the same energy, we use a rough estimate of $\epsilon=0.69$
obtained from an unpublished thermal-model calculation. Because this procedure 
does not model correlated nucleon coalescence or event-by-event cluster production, 
it should be interpreted only as a qualitative sensitivity test rather than a physical 
correction or a quantitative prediction.

Figure~\ref{fig:cumKap_sysModeHalfDeu} presents the cumulant ratios
($C_{2}/C_{1}$, $C_{3}/C_{1}$, and $C_{4}/C_{2}$) and factorial cumulant ratios
($\kappa_{2}/\kappa_{1}$, $\kappa_{3}/\kappa_{1}$, and
$\kappa_{4}/\kappa_{1}$) obtained with CIGAR in central collisions from the JAM
model at $\sqrt{s_{\rm NN}}=2.4$~GeV, for \(0.4<p_{T}<1.6\)~GeV/\(c\). The
abscissa is divided into four groups: (1) $-0.5<y<0.5$, w/ D; (2) $-0.5<y<0$,
w/ D; (3) $-0.5<y<0.5$, w/o D; and (4) $-0.5<y<0$, w/o D. Here ``w/ D''
denotes the original JAM results without light-nuclei formation, and ``w/o D''
denotes the results after applying the efficiency factor to mimic the removal
of free protons due to light-nuclei formation. For $C_{2}/C_{1}$,
$C_{3}/C_{1}$, $C_{4}/C_{2}$, and $\kappa_{2}/\kappa_{1}$, group~(1) gives the
lowest values, groups~(2) and~(3) are higher, and group~(4) gives the highest
values. Thus, both enlarging the rapidity window and not removing a fraction of the 
protons by forming deuterons will decrease these four ratios, qualitatively similar 
to the effect of reducing the rapidity window shown in Fig.~\ref{fig:cumKap_rap}. 
For $C_{4}/C_{2}$, the values in group~(1) are about $-0.8$ to $-0.5$, those in
group~(2) increase to about $-0.3$ to $-0.1$, those in group~(3) increase to
about $-0.2$ to $0.05$, with the two Au+Au results becoming positive, and those
in group~(4) reach about $0.3$ to $0.4$. The differences between Au+Au and
Ag+Ag and between cascade and mean-field modes are smallest in group~(4). For
$\kappa_{3}/\kappa_{1}$, group~(1) gives the largest values. For
$\kappa_{4}/\kappa_{1}$, all results are consistent with zero within
uncertainties, except for the Ag+Ag cascade result in group~(3), approximately $0.15$, which remains close to zero and does not imply any significant physical effect.

These results suggest that the negative $C_{4}/C_{2}$ values obtained without
light-nuclei formation may partly originate from the baryon-number-conservation
constraint, which is stronger for a larger acceptance and a larger number of
free protons. Reducing the rapidity window or removing a fraction of the
protons weakens this constraint and drives $C_{4}/C_{2}$ toward less negative
values. In addition, in group~(4), where both the acceptance and the number of
free protons are smallest, the differences between Au+Au and Ag+Ag and between
cascade and mean-field modes are visibly reduced. This indicates that at
smaller acceptance and lower proton multiplicity, the system-size and
mean-field effects are diluted and become less significant. A more realistic assessment of 
light-nucleus production should explicitly model deuteron and triton formation, for example 
through phase-space coalescence or a Wigner-function-based prescription, and propagate the 
associated correlations into the fluctuation observables; these improvements can be considered in future work.

\section{Summary}\label{sec:summary}

We have investigated proton-number cumulants and factorial cumulants up to fourth order in 
Au+Au and Ag+Ag collisions at \(\sqrt{s_{\mathrm{NN}}}=2.4\ \mathrm{GeV}\) with the JAM 
hadronic transport model, considering both cascade dynamics and a soft nuclear mean field. By 
employing the CIGAR framework, we have examined the effects of centrality determination and 
initial collision-volume fluctuations on these observables and established a simulation-based 
reference for studies at high baryon density.

Our analysis leads to several main findings. First, the cumulants obtained with CIGAR closely 
reproduce the \(N_{\mathrm{part}}\)-based CBWC benchmark, whereas RefMult3-based CBWC exhibits 
substantial biases at this low collision energy. Second, $C_{2}/C_{1}$, $C_{3}/C_{1}$, and $C_{4}/C_{2}$ 
generally decrease from peripheral to central collisions in both systems. The nuclear mean 
field modifies their magnitudes, with differences that depend on collision centrality and cumulant 
order. Third, increasing the rapidity acceptance suppresses these three ratios in the most central 
collisions, consistent with an enhanced sensitivity to global baryon-number conservation and 
other hadronic correlations. The fourth-order factorial-cumulant ratio $\kappa_{4}/\kappa_{1}$ remains 
consistent with zero within the present statistical precision for most of the investigated conditions.

We have also explored the possible influence of light-nucleus formation using a simplified 
proton-removal prescription. This test indicates that reducing the detected free-proton 
fraction shifts $C_{4}/C_{2}$ toward less negative values, similar to narrowing the rapidity 
window. Because the calculation does not explicitly simulate nuclear coalescence, the 
magnitude of this effect should not be interpreted quantitatively. Future work should 
incorporate realistic light-nucleus formation, assess the systematic robustness of the 
CIGAR reconstruction, and apply consistent analysis and acceptance procedures in direct 
comparisons with HADES data.

Taken together, the results provide a hadronic-transport baseline without an explicit QCD 
critical contribution and demonstrate that volume fluctuations, nuclear mean fields, 
baryon-number conservation, experimental acceptance, and free-proton depletion must be 
controlled when interpreting higher-order proton-number fluctuations at very low collision 
energies.

\section*{Acknowledgement}
This work is supported in part by the National Natural Science Foundation of China under Grant No. 12525509, No. 12447102, and the National Key Research and Development Program of China under contract No. 2022YFA1604900, and the Fundamental Research Funds for the Central Universities (XJ2026000302). Yu Zhang was supported by the Guangxi Natural Science Foundation (No. 2025GXNSFBA069067).

\bibliography{references}

\end{document}